# Hot-Carrier Seebeck Effect: Diffusion and Remote Detection of Hot Carriers in Graphene

*Juan F. Sierra[†,\*], Ingmar Neumann[†,‡], Marius V. Costache[†] and Sergio O. Valenzuela[†,¶,\*]*

† ICN2-Institut Catala de Nanociencia i Nanotecnologia, Campus UAB, Bellaterra, 08193 Barcelona, Spain.

‡ Universidad Autónoma de Barcelona, Bellaterra, 08193 Barcelona, Spain

¶ ICREA- Institució Catalana de Recerca i Estudis Avançats, 08010 Barcelona, Spain

ABSTRACT. We investigate hot carrier propagation across graphene using an electrical nonlocal injection/detection method. The device consists of a monolayer graphene flake contacted by multiple metal leads. Using two remote leads for electrical heating, we generate a carrier temperature gradient that results in a measurable thermoelectric voltage $V_{NL}$ across the remaining (detector) leads. Due to the nonlocal character of the measurement, $V_{NL}$ is exclusively due to the Seebeck effect. Remarkably, a departure from the ordinary relationship between Joule power $P$ and $V_{NL}$, $V_{NL} \sim P$, becomes readily apparent at low temperatures, representing a fingerprint of hot-carrier dominated thermoelectricity. By studying $V_{NL}$ as a function of bias, we directly determine the carrier temperature and the characteristic cooling length for hot-carrier propagation, which are key parameters for a variety of new applications that rely on hot-carrier transport.

Understanding the heating, energy flow and relaxation of two-dimensional carriers in graphene is essential for the design of graphene electronic devices.[1] In contrast to conventional metals with large Fermi surfaces, thermal decoupling of electrons from the crystal lattice leads to very slow electron-lattice cooling rates.[2-4] Electrons can be easily pushed out of thermal equilibrium with the lattice, even under weak electrical driving, and the generated hot carriers can propagate over extended distances.[5-7] Remarkably, such hot-carrier transport regime can occur at room temperature, resulting in novel thermoelectric and optoelectronic phenomena.[6] In combination with its large mobility and fast electrical response, these phenomena make graphene a material with great potential for a variety of applications, including bolometry, calorimetry and THz detectors.[8-10]

The inefficiency of carrier cooling originates from the intrinsic properties of graphene. Because of the large optical phonon energy $\hbar\Omega \sim 200$ meV, the most efficient mechanism available for hot-carrier cooling at low energies is the emission of acoustic phonons.[2-4] However, a small Fermi surface and momentum conservation severely restrict the acoustic phonons that can scatter off electrons. This leads to the observation of unconventional high-order cooling pathways assisted by disorder ("supercollisions", SC), which become dominant by relaxing the restrictions in phase space for acoustic phonon scattering.[11-13]

Experimentally, the temperature of hot carriers $T_C$ has been determined by means of spontaneous optical emission[14] and Johnson-noise thermometry measurements,[13,15] whereas their dynamics has been investigated using ultra-fast pump-probe spectroscopy.[16-20] However, the propagation of hot carriers injected by electrical driving has yet to be investigated. This is critical for the understanding of energy flow at the nanoscale and its control in high-speed devices.

In this work, we report hot-carrier propagation across monolayer graphene (MLG). Hot carriers are generated locally by a bias current, then diffuse away from the injection point, and are detected electrically with voltage probes in a region where no current circulates. The carrier temperature $T_C$ is deduced from the detected thermoelectric voltage $V_{NL}$ and the Seebeck coefficient $S$ of the sample. The temperature and current dependence of $V_{NL}$ enable us to identify when the carriers reach the detectors before thermalizing with the lattice, which remains at a lower temperature $T_L$. We demonstrate that the presence of hot carriers results in a strong increase of $V_{NL}$ at low bath temperatures $T_{bath}$ and a clear departure from the ordinary linear relationship between Joule power $P$ at the injector and $V_{NL}$, $V_{NL} \sim P$, which is typically found in conventional thermoelectric experiments for which $T_C \sim T_L$ (see Refs. 21, 22 and 23). Additionally, we observe that the bias dependence of $T_C$ is consistent with the energy relaxation rate predicted by the SC mechanism.[11] By measuring $V_{NL}$ as a function of the distance from the injection point, we can determine the characteristic cooling length $\xi$ for the electrically injected hot carriers. The measured voltage approaches 1 mV in our devices but we expect it to be much larger for high-quality graphene, resulting in a response that needs to be taken into account in high-frequency graphene transistors.

Graphene devices used in this work were prepared by mechanical exfoliation of MLG onto $p$-doped Si/SiO$_2$ (440 nm) substrates. The devices were prepared in two steps using electron-beam lithography. First, we deposited an amorphous carbon layer in the contact area just after exfoliation using electron-beam induced deposition (EBID), as described elsewhere.[24] Second, we defined electrical contacts with a width of 100 nm. The contacts were made by electron beam evaporation of Ti (5 nm)/Pd (60 nm) in a chamber with a base pressure of $10^{-8}$ Torr. The conducting $p$-Si substrate was used to apply a back-gate voltage $V_{BG}$ relative to the device to control the carrier density $n$ of the MLG. The presence of the amorphous carbon layer results in a contact resistance of ~5 kOhm, and fulfills the purpose of reducing the influence of the contacts in the hot-carrier dynamics. We have recently found that amorphous carbon does not affect the charge transport properties of graphene, preserving its mobility, but notably helps electrically detach the MLG from the leads.[24]

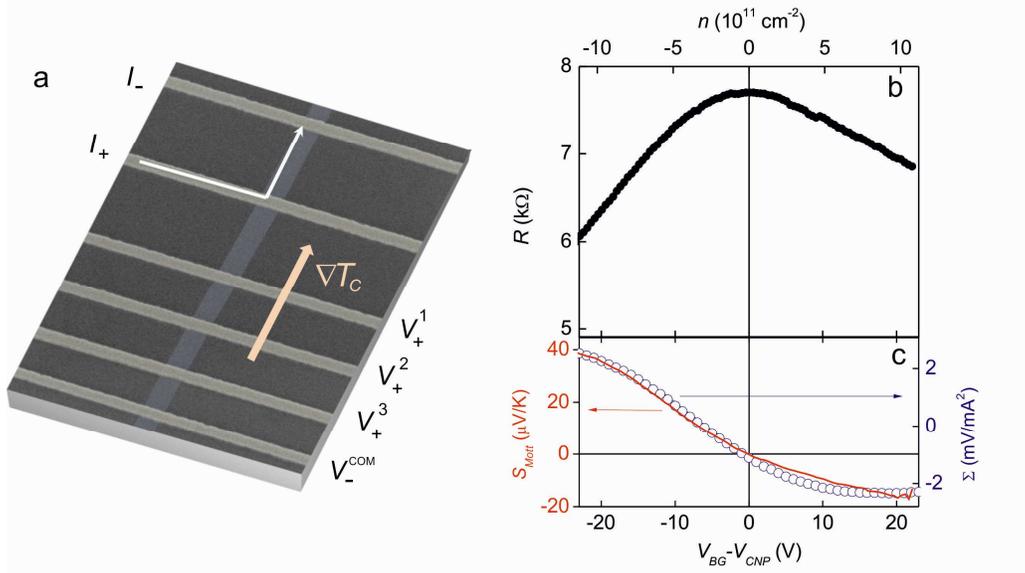

Figure 1. (a) Scanning electron microscope image (false color) of a monolayer graphene (MLG) device and the non-local measurement scheme. The electrical current $I$, applied between contacts $I_+$ and $I_-$, results in a thermal gradient $\nabla T_C$ along the device. The non-local signal is measured between contacts $V_+^d$ (being $d$ the detectors 1, 2 and 3) and $V_+^{COM}$. All the electrical contacts have a width of 100 nm. (b) Back-gate dependence of the square resistance in the detector 1 region at room temperature. The top axis shows the carrier density $n$. (c) Seebeck coefficient calculated from the Mott expression (left axis-solid line) and the quadratic fitting parameter $\Sigma$ (right axis-open circles) as a function of $V_{BG}$ at 296 K. We observe a good agreement between $S_{\mathrm{Mott}}$ and $\Sigma$ vs $V_{BG}$.

Figure 1(a) shows a scanning electron microscope image of one device with a schematic of the electrical connections. The measurement scheme mimics that of the nonlocal spin injection/detection technique that is commonly used to investigate spin transport.[25,26,27] A current $I$ between contacts $I_-$ and $I_+$ generates hot carriers that diffuse away from the injection region, which are then detected with remote voltage probes. The measured voltage $V_{NL}$ is said to be nonlocal (NL) because no electrical current flows in the detection region. In our devices, we define a set of three detectors with associated potential $V_+^d$ ($d$ = 1, 2 and 3) that are located at specific distances (1, 1.5 and 2 µm respectively) from the injector contact $I_+$ (see Fig. 1(a)). The thermal and hot-carrier transport properties are then investigated by measuring the nonlocal response $V_{NL}^d = V_+^d - V_-^{COM}$ in the remote detector $d$ relative to the common detector $V_-^{COM}$.

Samples are placed in a liquid helium continuous-flow cryostat that allows us to precisely control $T_{bath}$ between 10 K and 296 K. Measurements are carried out as a function of $T_{bath}$ and the back-gate voltage $V_{BG}$ in the linear $I$-$V$ regime, where scattering with interfacial $SiO_2$ substrate phonons and optical phonons is absent.[28-30] The three detectors allow us to determine the temperature of the carriers at different positions in the MLG. We have measured five devices that showed similar results. The data hereby presented were acquired with two of them (device 1 and 2). If not specified otherwise, all data shown correspond to detector 1.

We first realized a full electrical characterization of the devices. We carried out measurements where current is applied between the outermost contacts and the voltage measured between pairs of inner contacts. In device 1, graphene is slightly $p$-doped with the charge neutrality point (CNP) at $V_{CNP}$ = -2 V, the residual carrier density is $n_r = 10^{12}$ cm$^{-2}$ and the carrier mobility $\mu$ = 5,000 cm$^{-2}$/V·s at a carrier density of $n = 10^{12}$cm$^{-2}$. Figure 1(b) shows the gate-dependent square resistance $R$ measured in the first detector. We found that $R$ vs $V_{BG}$ presents a similar behavior independently of the voltage contacts selected, which demonstrates that the contact probes do not significantly modify the transport characteristics of the graphene underneath.

We then measured the non-local response $V_{NL}$ as a function of the applied current $I$ for different $V_{BG}$. Figure 2(a) shows measurements at room temperature ($T_{bath}$ = 296 K). At any given $V_{BG}$, we observe a parabolic dependence of $V_{NL}$ with current $I$. The parabolic behavior is verified in Fig. S1 of the Supporting Information, where it is observed that the data in Fig. 2(a) can be linearized by plotting $V_{NL}(I)$ vs $I^2$. The change from an upward ($V_{NL} > 0$) to a downward parabola ($V_{NL} < 0$) occurs progressively with $n$ and correlates with the change from electron to hole conduction; in particular we observe that $V_{NL}(I)$ is zero for all $I$ at the CNP ($n = 0$). Figure 2(b) shows $V_{NL}$ vs $n$ for specific applied currents (marked by vertical lines in Fig. 2(a)), where the progressive change from electrons to holes is clearly observed.

Notably, the parabolic dependence between $V_{NL}$ and $I$ (i.e, $V_{NL} \propto I^2$) morphs into $V_{NL} \propto I^{2\nu} \propto P^{\nu}$ with $\nu < 1$ for temperatures $T_{bath} < 100$ K (Fig. 2(c), see also Fig. S2). As for $T_{bath}$ = 296 K, $V_{NL}$ changes from positive to negative when passing from electron to hole conduction, with no signal at the CNP. However, $V_{NL}$ is larger and the change from upward to downward curvature is significantly more abrupt, as observed in $V_{NL}$ vs $n$ cuts at fixed $I$ (Fig. 2(d)).

To interpret the results in Fig. 2, we first note that carrier-carrier scattering processes are much faster than the electron-phonon scattering pathways,[31] independently of $T_{bath}$. As a consequence, a hot carrier population is established that can be described by a thermal distribution with a well-defined temperature $T_C > T_L$,[14,18,31] which decays away from the injector. Therefore, $V_{NL}^d$ can be expressed as

$$V_{NL}^d = - \int_{T_C^{COM}}^{T_C^d} S(T_C)\, dT_C \tag{1}$$

where $T_C^d$ is the carrier temperature at the detector $d$ and $T_C^{COM}$ is the carrier temperature at the common detector. As discussed below, $T_C^{COM} \sim T_L \sim T_{bath}$ because the distance between the injector and the common detector (2.5 µm) is much larger than the cooling length $\xi$.

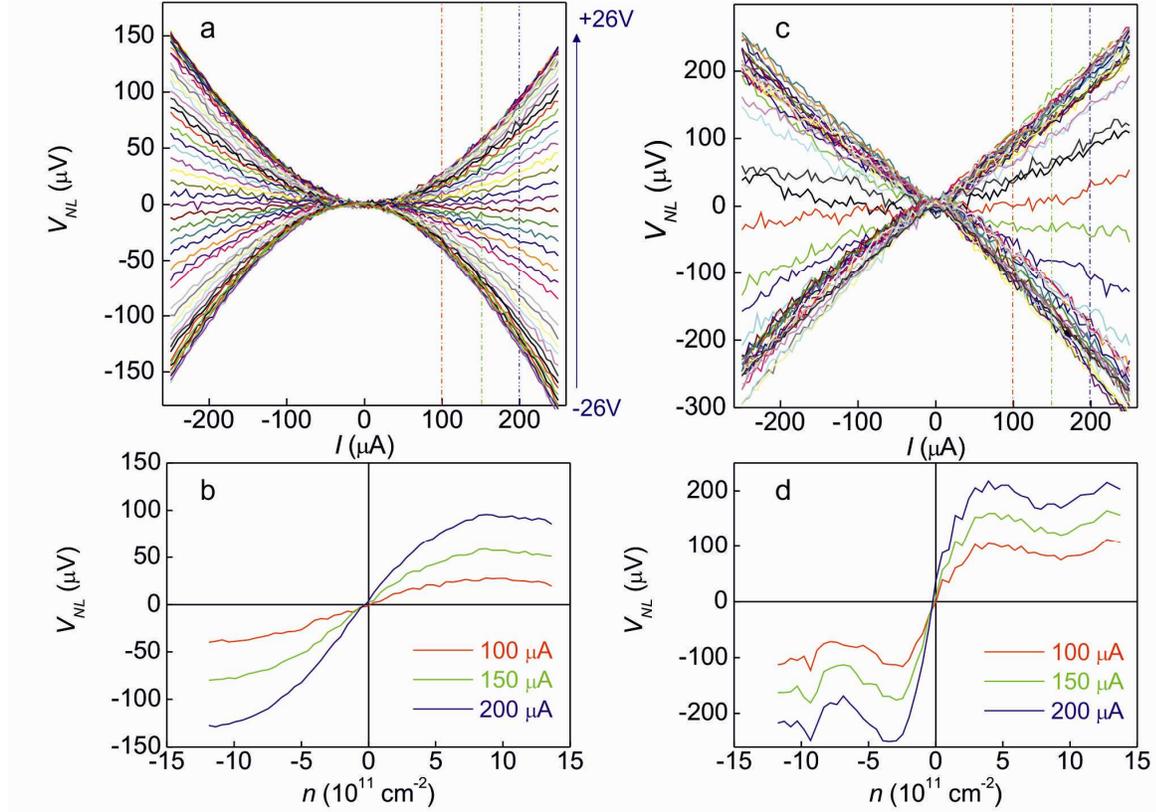

Figure 2. (a) $V_{NL}$ vs $I$ for different $V_{BG}$ (from -26 V ($n = -13 \times 10^{11}$ cm$^{-2}$) to 26 V ($n = 13 \times 10^{11}$ cm$^{-2}$)) and (b) $V_{NL}$ vs $n$ at fixed $I$ (100 µA, 150 µA, 200 µA). Measurements in (a) and (b) are performed at 296 K. (c) and (d) show analogous measurements at 10 K. At 296 K, a quadratic dependence of $V_{NL}$ vs $I$ is observed. At 10 K, the presence of hot carriers in the detectors results in a larger $V_{NL}$ and a strong departure from such quadratic dependence.

We also note that, as shown in Refs. (21, 22 and 23), the thermopower in graphene can be predicted by the semiclassical Mott relation,[32]

$$S_{\text{Mott}} = \left(\frac{\pi^2 k_B^2 T_C}{3e}\right)\left(\frac{\partial \ln R}{\partial E}\right)_{E=E_F} \quad (2)$$

where $e$ is the electron charge, $k_B$ is the Boltzmann constant, and $E_F$ is the Fermi energy. $S_{\text{Mott}}$ is therefore a good approximation for the graphene Seebeck coefficient $S$ in Eq. (1). The calculated $S_{\text{Mott}}$ at 296 K is shown in Fig. 1(c). $S_{\text{Mott}}$ reaches a maximum value of 40 µV/K and changes from positive to negative at the CNP, indicating the nature of the majority charge carriers in the system.

The parabolic behavior found in Fig. 2(a) can be understood from the Joule dissipation at the current injector and the cooling rates involved. At large enough bias, heat diffusion into the leads and into the unbiased graphene regions can be neglected and thus the cooling of the carriers is mediated by the electron-lattice coupling.[15] In addition, at room temperature, we expect that the carrier cooling time for our MLG is ~10 ps,[11] and the typical cooling length $\xi < 100$ nm, which is much smaller than the electrode separation. Therefore, the carriers are thermalized with the lattice ($T_C = T_L$) at the detector region all the way to within 100 nm of the injector.

Under these conditions,[15] the steady state solution of the heat equation shows that the temperature gradient $\nabla T_C$ at the detectors is $\nabla T_C \propto I^2 \propto P$. For $T_C^d - T_C^{\text{COM}} \sim T_C^d - T_{bath} \ll T_{bath}$, we obtain, using Eq. (1), $V_{NL}^d \sim -S(T_{bath})(T_C^d - T_{bath}) \propto -S(T_{bath}) I^2$, that is, $V_{NL}^d \propto P$. Therefore, at sufficiently high $T_{bath}$, the thermoelectric response of our devices is similar to that found using an external heater,[21-23] despite the fact that, in our case, graphene is part of said heater.

In contrast, at low enough temperatures (below ~100 K), the hot-carrier lattice thermalization length can exceed a few hundred nanometres[11] and, therefore, the conditions $T_C = T_L$ and $T_C^d - T_{bath} << T_{bath}$ are no longer satisfied at the detectors. As illustrated in Fig. 2(c) for $T_{bath}$ = 10 K, this leads to a remarkable increase in $V_{NL}$ and to a dramatic departure from the linear relationship between $V_{NL}$ and $P$. Such behavior contrasts to that observed in conventional thermoelectric experiments, where the relationship $V_{NL} \propto P$ is valid at all temperatures.[21-23] This is verified in Fig. S3, where we show thermoelectric measurements using an external heater. In those conventional experiments, the temperature pre-factor in Eq. (2) reduces the thermoelectric voltage by a factor ~30 from room temperature to 10 K. In our hot-carrier Seebeck measurements, we observe that $V_{NL}$ is actually larger at $T_{bath}$ = 10 K than at $T_{bath}$ = 296 K. By inspecting Eq. (1) and (2), this is possible for $T_C^d \approx 100$ K.

The presence of hot carriers and the main relaxation by disorder-mediated scattering, or supercollisions (SC), accounts for the change in the functional dependence at low temperatures. Recent experiments using Johnson-noise thermometry in two-terminal devices[13] have demonstrated that $T_C$ can reach 400-700 K with a Joule power $P = 0.2$ mW/μm². Therefore, assuming $T_C^d \gg T_L; T_{bath}$ and $T_C^{COM} \approx T_{bath}$, we integrate Eq. (1). By taking into account Eq. (2), we obtain $V_{NL}^d \propto T_C^2$, which combined with the predicted supercollision energy power loss $A(T_C^3 - T_L^3)$, results in $V_{NL}^d \propto P^{2/3}$, and thus $\nu$ = 2/3, which is in agreement with our experiments.

The increase that we observe in $V_{NL}$ at low temperatures is also a signature of SCs, as the dependence expected from momentum-conserving scattering by acoustic phonons predicts the opposite behavior.[11] This phenomenon can be understood by considering the temperature dependence of the cooling rate $\gamma$ for hot carriers in the regime $T_L \geq T_{BG}$, where $T_{BG}$ is the Bloch-Grüneisen temperature, which in graphene sets the boundary between direct electron-phonon scattering ($T_L < T_{BG}$) and the regime which is dominated by SCs ($T_L \geq T_{BG}$). Direct emissions of acoustic phonons are rare leading to a cooling rate $\gamma_{e-p} \propto 1/T_L$ that would produce a decrease of the non-local response with decreasing temperature.[2,3] However, SCs lead to a cooling rate $\gamma_{SC} \propto T_L$ where the proportionality constant is related with the amount of disorder in the system.[11] The decrease of the cooling rate with decreasing temperature in the latter model explains the increase in the nonlocal response.

The nonlocal voltage decreases monotonically with temperature, indicating that we do not achieve the regime for direct emissions of phonons. The reason is twofold. First, the high disorder concentration in our devices hinders the direct emission of phonons, and second, the injection of current likely results in $T_L > T_{BG}$, even for $T_{bath} < T_{BG}$. Indeed, the expression for $T_{BG}$ is given by $k_B T_{BG} = \left(2v_s/v_F\right) E_F$, with $v_s$ and $v_F$ the sound and the Fermi velocity respectively. This temperature can be tuned with $E_F$, i.e with the back-gate voltage. For the maximum back-gate voltage applied to our samples (26 V), we obtain $T_{BG}^{max} = 20$ K. This means that $T_{BG}^{max}$ is not far from the lowest temperatures that we achieve in our experiments and, therefore, it might be difficult to access the regime where $T_L < T_{BG}$, except perhaps at low currents. This argument is supported by the data in Fig. 2c, which appears to be somewhat more parabolic at low $I$ for large gate voltages, although the results are not conclusive (see Fig. S4).

The abrupt change in the $V_{NL}$ vs $n$ cuts at fixed $I$ (Fig. 2(d)) is a consequence of a larger rise in $T_C$ close to the CNP. This is partly due to the fact that the Joule power dissipated in the MLG is distributed over a relatively small number of carriers close to the CNP. Therefore, even though the overall shape of the Seebeck coefficient vs back-gate voltage is unchanged with temperature, the temperature gradient at the detector is gate dependent at low temperatures. This does not occur at room temperature because the dissipation, and temperature gradient, is dominated by the high-resistance contacts.

The same result is observed in the other measured devices. Figure 3 shows a comparison for device 2 between $S_{Mott}$ calculated at 296 K and at 77 K and $V_{NL}$ measured at the same temperatures. For this device, $V_{NL}$ has a similar magnitude at 77 K and at 296 K. However, the shape difference between them close to the CNP, due to the larger rise in $T_C$, is evident; such a difference is not observed in $S_{Mott}$ (inset Fig. 3).

For a quantitative understanding of the thermoelectric response, we fit our results in Fig. 2 (a) to $V_{NL} = V_l + \Sigma I^2$, with $V_l$ and $\Sigma$ as the fitting parameters. Here, $\Sigma \propto S_G - S_l$, with $S_G$ the graphene Seebeck coefficient; $V_l$ and $S_l$ account for a small gate-independent thermoelectric voltage generated along the measurement lines. The resulting $\Sigma$ vs $V_{BG}$ is shown in Fig. 1(c) (open symbols). As discussed above, $S_{Mott}$ predicts the thermopower in graphene[21-23]; we thus expect that $\Sigma$ will closely follow $S_{Mott}$. Indeed, the comparison between $\Sigma$ and $S_{Mott}$ shows excellent agreement. Moreover, from the proportionality constant between $\Sigma$ and $S_{Mott}$, we calculate that $T_C^d - T_{bath} \sim 83$ K/mA$^2$ for detector 1. At $I = 250$ μA, this is equivalent to about 5 K, which proves that $T_C^d - T_{bath} \ll T_{bath}$. Such values agree with those estimated with numerical simulations using COMSOL and realistic device parameters.

For intermediate temperatures (< 150 K), we found that the experimental results can be fitted to the empirical expression $V_{NL} = V_l + \beta |I|^{2\nu} + \Sigma(T_{bath})I^2$, with $\nu = 2/3$. Here, $V_l$ and $\beta$ are the only fitting parameters, $|I|^{4/3}$ stands for the functional dependence $V_{NL}^d \propto P^{2/3}$, and $\Sigma(T_{bath}) = \Sigma(296 \text{ K})(\frac{T_{bath}}{296 \text{ K}})$ (see Eq. (2)). We observe that $\beta$, which in a way represents the thermoelectric voltage due to the excess temperature of hot carriers, increases rapidly at low temperatures, especially below 100 K.

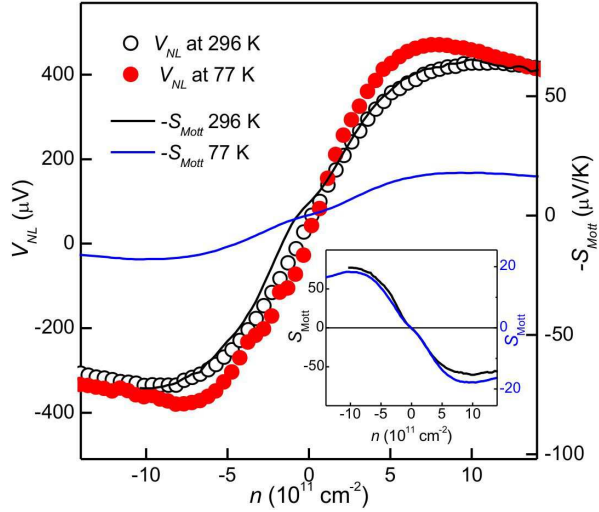

Figure 3. $V_{NL}$ vs $n$ measured at 296 K (open symbols) and 77 K (full symbols) as well as the calculated Seebeck coefficient from the Mott expression (line) at the same temperatures (device 2). At 296 K, $V_{NL}$ is well described by $S_{Mott}$ while at 77 K, deviations near the CNP ($n = 0$) are observed due to hot carriers. Inset: $S_{Mott}$ at 296 K (left axis) and at 77 K (right axis).

For an additional test of the SC mechanism, we extract $T_C$ as a function of the dissipated power $P$ in MLG.[33] We integrate Eq. (1) and take into account Eq. (2), obtaining,

$$T_C = \sqrt{(T_L)^2 + 2\left(\left|V_{NL}/\kappa\right|\right)}, \qquad (3)$$

where $\kappa = \left(\frac{\pi^2 k_B^2}{3e}\right)\left(\frac{\partial \ln R}{\partial E}\right)_{E=E_F}$. For $T_{bath} = 10$ K, we estimate that the carrier temperature at the first voltage detector is as high as $T_C = 165$ K for $P \sim 0.65$ mW/μm$^2$ and $n \sim 2.5 \cdot 10^{11}$ cm$^{-2}$ (see inset in Fig. 4(a)). Figure 4(a) illustrates $T_C^3/P$ vs $P$. As predicted for the SC model, a plateau ($P \propto T_C^3$) is evident for sufficient Joule power $P$ where the condition $T_L \geq T_{BG}$ is fulfilled, whereas $T_C$ rises faster nearby the CNP due to the relatively small density of carriers, as discussed above.

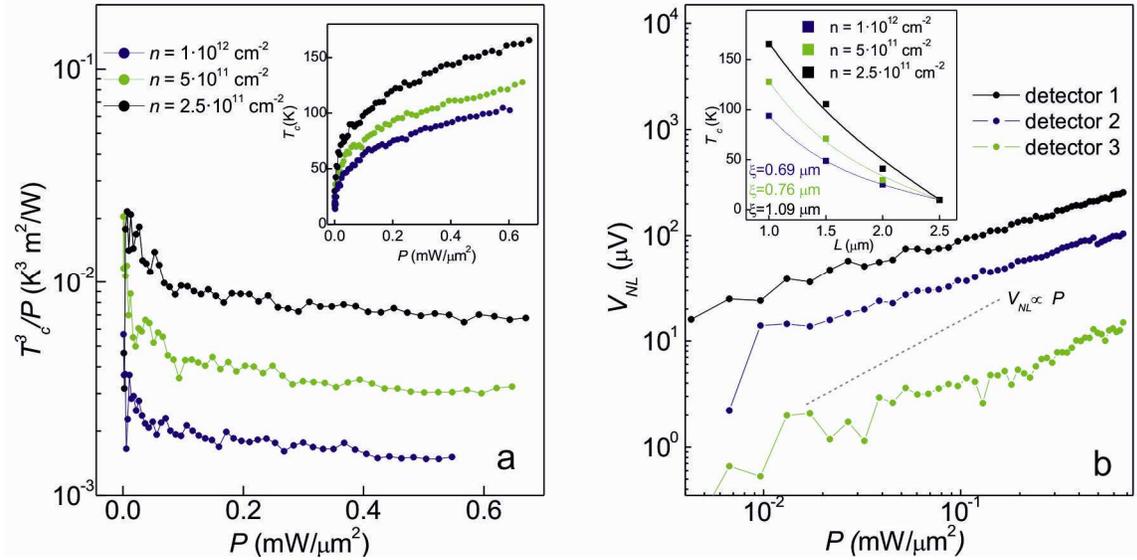

Figure 4. (a) (Inset) Carrier temperature $T_C$ at detector 1 as a function of the Joule power $P$ (per unit area) that is dissipated in the MLG for the indicated carrier densities, $n$. In the representation $T_C^3/P$ (main figure) a plateau develops. The carrier temperature is larger as $n$ approaches the CNP, mostly due to the reduced concentration of carriers. (b) $V_{NL}$ vs $P$ (log-log scale) measured in the three detectors. Dash-grey line indicates slope 1 ($V \propto P$). Inset: $T_C$ vs distance $L$ from the injector for the indicated $n$. The lines are the best fits to the steady state temperature profile expression in the main text. Data in (a) and (b) are measured at 10 K.

Finally, we use the three detectors (Fig. 1 (a)) to estimate the cooling length $\xi$. At $T_{bath} = 10$ K we measure the same sub-linear dependence of $V_{NL}$ vs $P$ in the three detectors, indicating the presence of hot carriers in all of them. This is demonstrated in Fig. 4(b), where $V_{NL} \propto P$ is shown for comparison. The temperature profile across the detectors is evaluated using Eq. (3) for $P \sim 0.65$ mW/µm$^2$, presenting a characteristic cooling length on the order of 1 µm. Because of the large variations of $T_C$, it is unclear if the profile obeys an exponential or a power law. In photo-thermoelectric experiments,[5,7] $\xi$ was obtained using the heat diffusion equation whose general solution is $T_C(x) - T_L \propto \sinh\left[(D-x)/\xi\right]$, where $D = 2.5$ µm is the distance between the injector and the common voltage detector. By fitting our data with this model, we obtain $\xi = 0.7$-$1.1$ µm for $n$ in the range of $1 \times 10^{12}$ cm$^{-2}$ to $2.5 \times 10^{11}$ cm$^{-2}$ (see inset of Fig. 4(b)).

Our devices show a rather fast relaxation of hot carriers. This is likely related to a relatively high concentration of impurities, which results in large SC rates. Extrapolation to $x = 0$ leads to hot-carrier temperatures of 370-450 K at the injector. These values are lower than those reported in Ref. 13, within a factor of 2, which might be related to a lower density of scattering centers in graphene on boron-nitride substrates.

In summary, we have proposed and implemented a new experimental technique based on non-local measurements to detect the presence of hot carriers in graphene. The technique allows us to directly determine the carrier temperature in remote voltage probes. Analysis of the non-local response as a function of temperature shows two different regimes. At high temperatures, the magnitude of the detected signal and its linear relationship with the dissipated Joule power can be accounted for by a model where the electrons are thermalized with the lattice. In essence, the results do not differ from those obtained with an electrically isolated heater. For low temperatures, however, the non-local response presents a clear departure from the above linear relationship, with a signal much larger than that expected by simple Joule heating of the lattice. We demonstrate that the magnitude of the signal and its functional dependence with power are strong evidence of hot-carrier generated thermoelectric voltages and that the supercollision mechanism is the predominant cooling pathway for hot-carrier cooling. Future experiments could investigate the relative weight of the different electron-phonon scattering processes in multilayer graphene, in particular at high bias.[34]

Beyond carrier-phonon physics and novel hot-carrier thermoelectricity phenomena, our work has important implications for the design of high-speed graphene-based devices. In the low temperature regime, the magnitude of the hot-carrier thermoelectric signal is as large as a few hundred microvolts but it can be much larger in high-quality graphene. Because the peak Seebeck coefficient scales with $1/\sqrt{n_r}$, a decrease in the residual carrier concentration $n_r$ by two orders of magnitude, which is typical for suspended graphene or graphene on boron nitride, will result in a tenfold increase in the signal. The signal will be further enhanced by an increase of the mean-free path, which will decrease the rate of supercollisions as well as strongly increase the carrier mobility and diffusion constant. Therefore, the hot carriers will be longer lived and diffuse much further than in our devices. Under these conditions, it is plausible that the signal can be as large as a few hundred mV, even at room temperature. It will thus strongly impact the performance of conventional graphene devices, and at the same time, create new opportunities for nanoscale bolometry and calorimetry.

## AUTHOR INFORMATION

**Corresponding Author**

* juan.sierra@icn.cat, SOV@icrea.cat

ACKNOWLEDGMENTS

We thank B. Raes, F. Bonell and A. W. Cummings for a critical reading of the manuscript. We acknowledge the support from the European Research Council (ERC Grant Agreement No. 308023 SPINBOUND), the Spanish Ministry of Economy and Competitiveness, MINECO (under contracts MAT2013-46785-P and Severo Ochoa Program SEV-2013-0295), and the Secretariat for Universities and Research, Knowledge Department of Generalitat de Catalunya. JFS acknowledges support from the Juan de la Cierva (JCI-2012-12661) and the Beatriu de Pinós programs, and MVC from the Ramón y Cajal program (RYC-2011-08319).

Supporting information for

# Hot-Carrier Seebeck Effect: Diffusion and Remote Detection of Hot Carriers in Graphene

*Juan F. Sierra[†,\*], Ingmar Neumann[†,‡], Marius V. Costache[†] and Sergio O. Valenzuela[†,¶,\*]*

† ICN2-Institut Catala de Nanociencia i Nanotecnologia, Campus UAB, Bellaterra, 08193 Barcelona, Spain.

‡ Universidad Autónoma de Barcelona, Bellaterra, 08193 Barcelona, Spain

¶ ICREA- Institució Catalana de Recerca i Estudis Avançats, 08010 Barcelona, Spain

**Content**

Linearized data in Fig. 2(a) by plotting the nonlocal voltage $V_{NL}$ vs $I^2$ in order to help visually verify the parabolic dependence in this figure; $V_{NL}$ vs $I$ for selected temperatures; measurements using external heater and comparison with results using device 1; $V_{NL}$ vs $I$ for low $I$ at 10 K, showing a parabolic-like behavior.

**Corresponding Author**

\* juan.sierra@icn.cat, SOV@icrea.cat

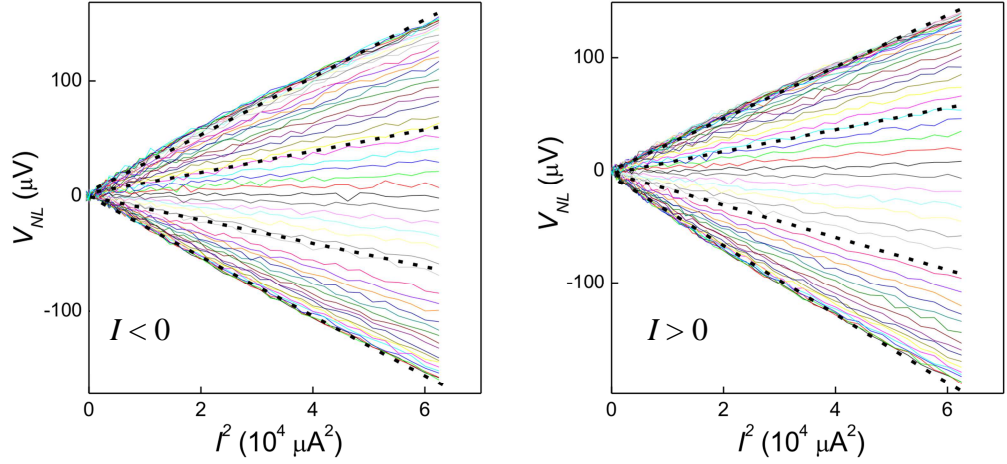

Figure S1. Linearized data from Fig. 2(a) by plotting the nonlocal voltage $V_{NL}$ vs $I^2$. This helps verify visually the parabolic dependence in this Fig. 2(a). Measurements for negative and positive currents, $I$, are shown in the left and right panels, respectively. The black dashed-lines are guides to the eye.

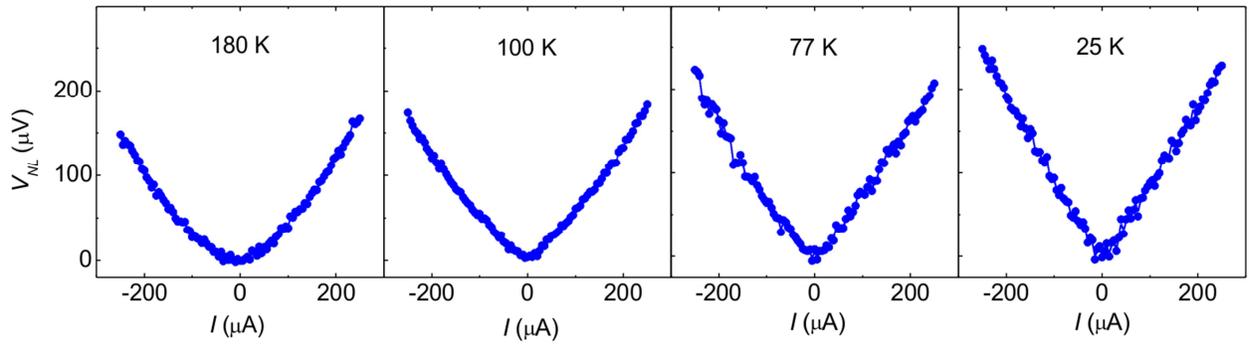

Figure S2. $V_{NL}$ vs $I$ for selected temperatures. At $T_{bath}$ = 180 K, $V_{NL}$ is parabolic; at $T_{bath}$ = 100 K and below, $V_{NL}$ is larger and markedly non-parabolic.

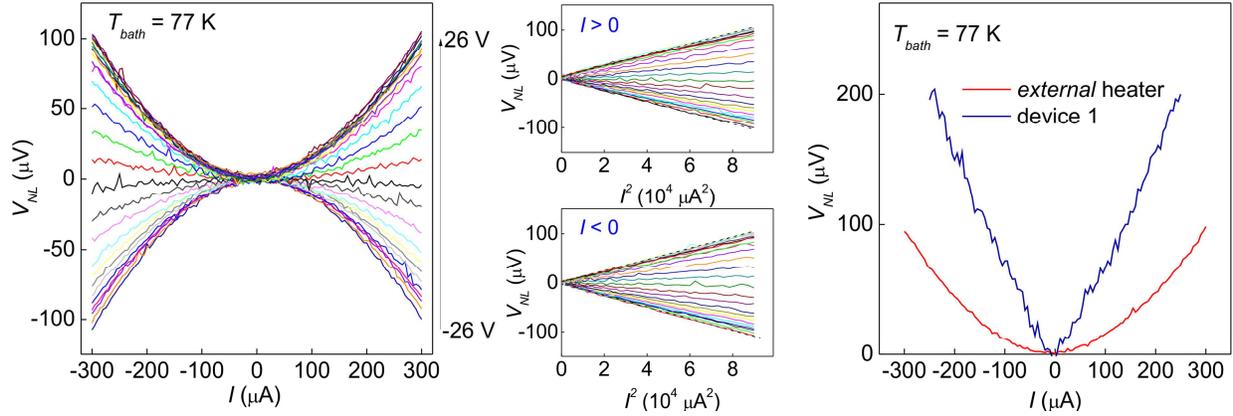

Figure S3. Left panel: Non-local voltage, $V_{NL}$ vs $I$ at $T_{bath}$ = 77 K for different back-gate voltages in a sample with an external heater. Middle panels: linearized data, $V_{NL}$ vs $I^2$, from the left panel for negative (bottom) and positive (top) currents $I$. Right panel: $V_{NL}$ vs $I$ at $T_{bath}$ = 77 K for device 1 (blue) and for the device with external heater (red); $n = 1 \times 10^{12}$ cm$^{-2}$.

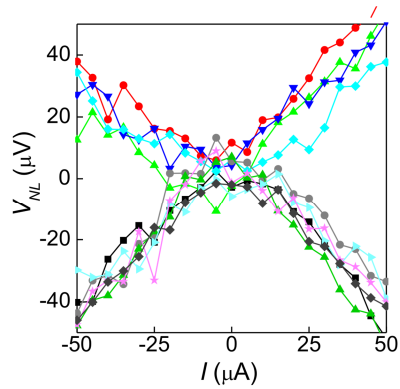

Figure S4. $V_{NL}$ vs $I$ for low $I$ at 10 K, showing a parabolic-like behavior only at very low currents ($I < 25$ μA) and more clearly at negative back-gate voltages (negative $V_{NL}$). The results are non-conclusive due to the noise level of the measurements.